\documentclass[aps,prl,amsmath,amssymb,amsfonts,twocolumn,
superscriptaddress,floatfix,footinbib,showpacs]{revtex4}

\usepackage{graphicx}

\newcommand{\vv}{ {\bf v}}
\newcommand{\xx}{ {\bf x}}
\newcommand{\kk}{ {\bf k}}
\newcommand{\eq}[1]{(\ref{eq:#1})}

\begin{document}
\title{Bogoliubov-{\u C}erenkov radiation in a
  Bose-Einstein condensate\\ flowing against an obstacle}

 \author{I. Carusotto}
  \affiliation{CNR-BEC-INFM, Trento, I-38050 Povo, Italy}

 \author{S.X. Hu }
\altaffiliation{Present address: Laboratory for Laser Energetics,
University of Rochester, 250 E. River Road, Rochester, NY 14623}
 \affiliation{Los Alamos National Laboratory, Los Alamos, 87544 NM}

 \author{L. A. Collins}
 \affiliation{Los Alamos National Laboratory, Los Alamos, 87544 NM}

 \author{A. Smerzi}
 \affiliation{Los Alamos National Laboratory, Los Alamos, 87544 NM} 
  \affiliation{CNR-BEC-INFM, Trento, I-38050 Povo, Italy}

\begin{abstract}
We study the density modulation that appears in a Bose-Einstein
condensate flowing with su\-per\-so\-nic velocity against an obstacle. 
The experimental density profiles observed at JILA are reproduced
  by a numerical integration of the Gross-Pitaevskii equation and
 then interpreted in terms of \u Cerenkov emission of Bogoliubov
excitations by the defect.
The phonon and the single-particle regions of the Bogoliubov spectrum
are respectively responsible for a conical wavefront and a fan-shaped
series of precursors.
\end{abstract}

% \date{\today}

 \maketitle

%{\it Introduction}. 
The \u Cerenkov effect was first discovered in the electromagnetic
radiation emitted by charged particles traveling 
through a dielectric medium at a speed larger than the medium's phase
velocity~\cite{LandauECM}. 
A charge moving at the speed $\vv$ is in fact able to resonantly
excite those modes of the electromagnetic field which satisfy the kinematic \u Cerenkov
resonance condition $\omega_{em}(\kk)=\vv\cdot \kk$: part of the
kinetic energy of the particle is then emitted as \u Cerenkov
radiation, with a peculiar frequency and angular spectrum~\cite{jelley}.
Electromagnetic waves in a non-dispersive medium of refractive index $n$ have a
linear dispersion law relation $\omega_{em}(k)= c k /n$: the \u
Cerenkov condition is then satisfied on a conical surface in $k$-space
of aperture $\cos\phi=c/(n v)$, which corresponds to a conical 
wavefront of aperture $\theta=\pi/2-\phi$ behind the particle.
Thanks to the interplay of interference and propagation, much richer
features appear in the spatial and $k$-space pattern of \u Cerenkov
radiation in dispersive media~\cite{Afanasiev,carusotto01} and
photonic crystals~\cite{Cerenk_PC}.

The concept of \u Cerenkov radiation can be generalized to any system
where a source is uniformly moving through a homogeneous medium at a
speed larger than the phase velocity of some elementary excitation to
which the source couples.
Many systems have been investigated in this perspective, ranging from
e.m. waves emitted by the localized nonlinear polarization induced by
a strong light pulse travelling in a nonlinear
medium~\cite{fs_cherenk,dipolar_cherenk}, to the sonic waves 
generated by an airplane moving at supersonic velocities, to phonons
in a polaritonic superfluid~\cite{Polar_superfl}, and in a broader
sense, to the surface waves emitted by a boat moving on the quiet
surface of a lake~\cite{lake}.
In this Letter we compare our theoretical results of the density
perturbation induced in a Bose-Einstein condensate (BEC) which flows
against a localized obstacle at rest with the experimental images
taken by the JILA group~\cite{pc}. 
Modulo a Galilean transformation, the physics of a moving source in a
stationary medium is in fact equivalent to the one of a uniformly
moving medium interacting with a stationary defect.
The experiment has been performed by letting a BEC expand at
hypersonic speed against the localized optical potential 
of a far-detuned laser beam. The observed density profiles are
successfully reproduced by numerically solving the time-dependent
Gross-Pitaevskii equation and physically interpreted by a simple
model of \u Cerenkov emission of Bogoliubov excitations by a weak
defect.
 \begin{figure}[htbp]
    \begin{center}
  \includegraphics[width=3.2cm,clip]{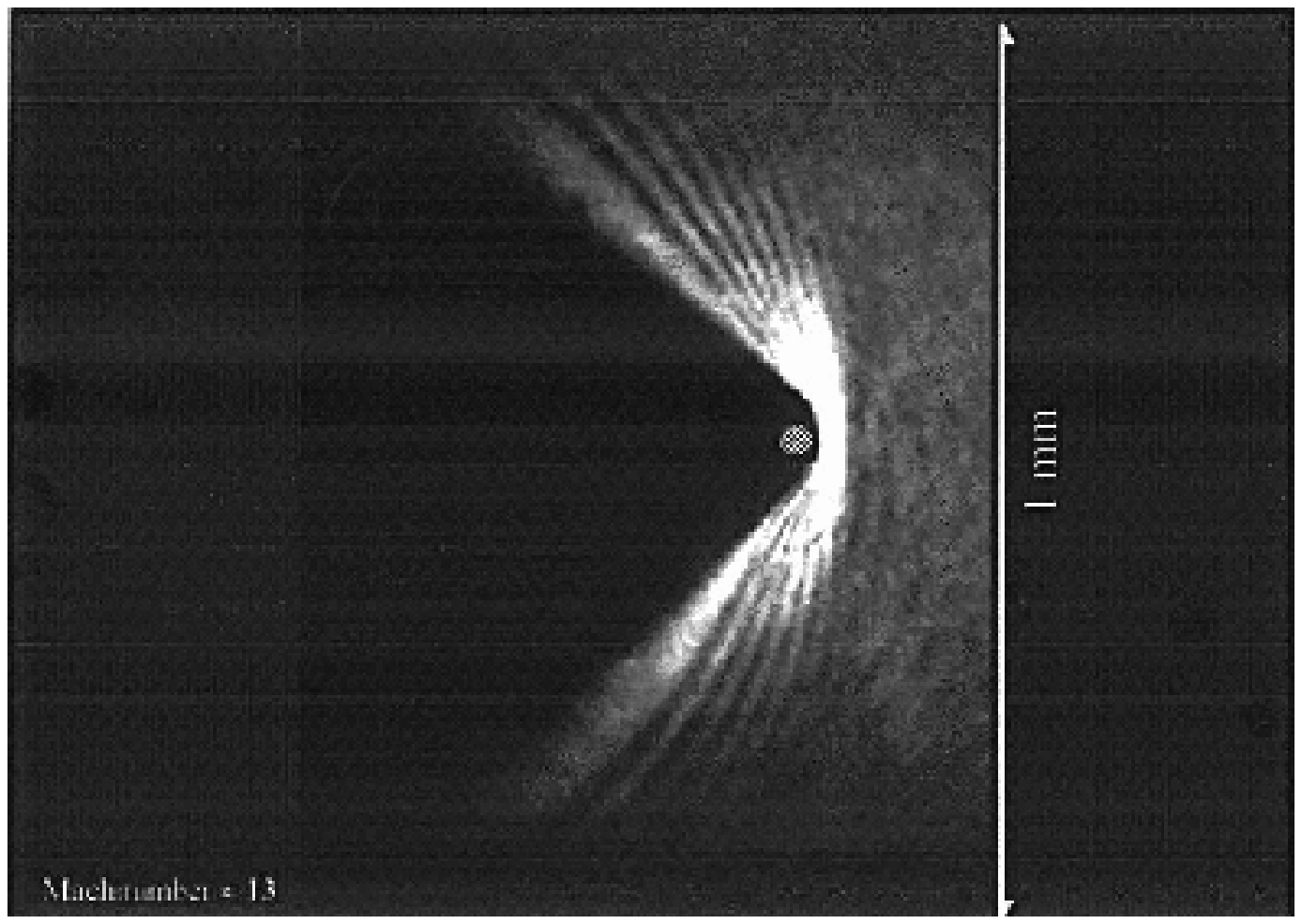} 
\hspace{0.5cm}
\includegraphics[width=3.2cm,clip]{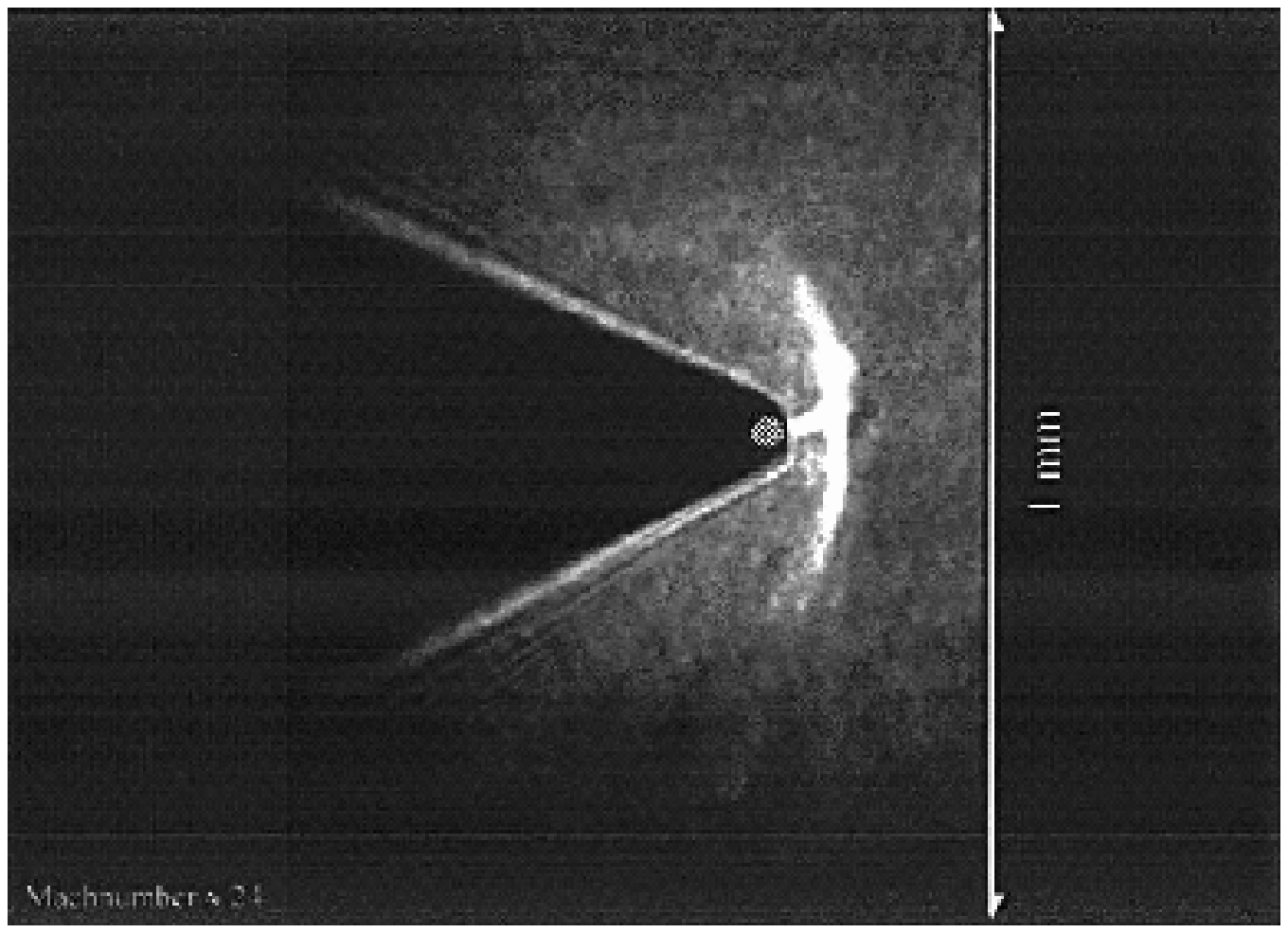}
      \caption{Experimental~\cite{pc} density profiles (integrated
	along $z$) 
of a BEC hitting an obstacle at supersonic velocities
$v/c_s= 13$ (a) and $24$ (b). The angles of the conical
wavefronts are $sin(\theta) = 0.73$ and $sin(\theta) = 0.43$, respectively.
The condensate flow is from the right to the left.}
 \label{Exp}
    \end{center}
  \end{figure}

{\it The experiment}. The experimental results analyzed in this paper
have been obtained by the JILA group~\cite{pc} with 
a gas of $N=3 \times 10^6$  Bose-Einstein
condensed $^{87}\textrm{Rb}$ atoms confined in 
a cylindrically symmetric harmonic trap of frequencies $\{\omega_r,
\omega_z \} =2 \pi \{ 8.3, 5.3 \}\,\textrm{Hz}$.
The BEC is slightly cigar shaped, with the long axis pointing in the
direction  of gravity ($z$-axis), and a Thomas-Fermi radius in the
$x-y$ plane of $31.6~\mu$m.
At $t = 0$, the sign of the confining potential in the $xy$ plane is
inverted, so that the BEC undergoes an antitrapped expansion in the
$xy$ plane. At the same time, the confinement along $z$ is switched
off and replaced by a linear magnetic field gradient to cancel the
effect of gravity. 
The obstacle consists of the repulsive potential of a blue-detuned
dipole beam with a round Gaussian profile.
%The laser power is $4.9$ mW, the light wave-length $660$ nm and the
%Gaussian beam waist $17$ $\mu$m.  
The beam is directed in the $z$-direction and placed in the
vicinity of the trapped condensate, and remains in this position during
the experiment.
Images of the BEC density profile after different expansion times
$t_{exp}$ are then taken by means of destructive absorption
imaging. Two examples are shown in Fig.(\ref{Exp}).
The field of view is centered in the region around the defect in
order to observe the perturbation that it creates in the expanding
condensate. In this region, the condensate is
flowing from the right to the left.
During the expansion, the condensate is strongly accelerated by the
inverted harmonic trap, so that the local density at the defect
position decreases in time, and the local speed $v$ increases.
As a result, images taken after increasing expansion times correspond 
to increasing values of the Mach number $v/c_s$, $c_s$ being the local
value of the sound velocity at the defect position.
In Fig.\ref{Exp}(a,b), the expansion times are 
$t=33.6,\,50\,\textrm{ms}$, long enough for the flow to be strongly
supersonic $v/c_s=13,\,24$.
Downstream of the defect the atomic density is strongly reduced in a
conical shadow region, separated from the unperturbed condensate
region by a conical wavefront followed by a fan-shaped series of
 precursors which extends far in the upstream direction.
For increasing Mach number $v/c_s$, the aperture of the shadow area
decreases, as well as the spatial period of the precursors.
Physical interpretations for all these features will be provided in
the following.

\begin{figure}[htbp]
    \begin{center}
\includegraphics[width=7.cm,clip]{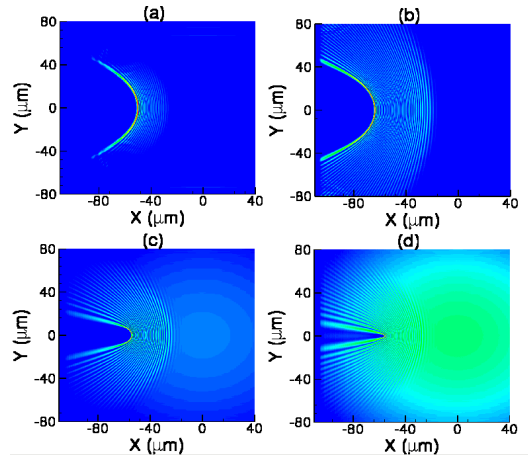}
      \caption{GPE density profiles integrated along the $z$-direction
obtained with different expansion times and sizes of the obstacle :
a) $t=33.6$ms, w=$17~\mu$m.; b) $t=50$ms, w=$17~\mu$m;
c) $t=38.4$ms, w=$3.74~\mu$m; d)  $t=38.4$ms,
w=$0.374~\mu$m. } 
  \label{GPE_sim}
    \end{center}
  \end{figure}

{\it Gross-Pitaevskii simulations}.
As a first step, the experimental data have been reproduced by
a numerical simulation of the 3-dimensional, time-dependent Gross-Pitaevskii
equation by means of a real-space-product finite element
discrete-variable-representation~\cite{collins98}, a numerical method 
that has already seen effective service in solving several large-scale
computational problems involving BECs~\cite{denschlag00}. 
Results for the exact parameters of the experiment are shown in
Fig.\ref{GPE_sim}(a,b).

The laser obstacle consists of a repulsive potential with a Gaussian
shape $V(r)= V_{max}\,\exp(-2 r^2/w^2)$.
The maximum height $V_{max}= 4961\,\hbar\omega_r$ of the potential is
much larger than the chemical potential of the BEC in the initial
trapped configuration and its width $w=17\,\mu\textrm{m}$
is $\sim 50$ times larger than the BEC healing length: the obstacle
can therefore 
considered, for all practical purposes, as a rigid cylinder.
In Figs.\ref{GPE_sim}(a,b), the defect is situated at a distance of
respectively $d=75,\,87\,\mu\textrm{m}$ from the condensate center and
the imaging times are $t=33.6,\,50\,\textrm{ms}$.
The density profiles emerging from the simulations are in agreement
with the experimental findings of Fig.(\ref{Exp}).
The Mach numbers in the GPE simulations are $12.4$
(a) and $23.8$ (b), and the angles of the
conical wavefronts delimiting the shadow areas are 
$\sin(\theta) = 0.7$ (a) and $sin(\theta) = 0.48$ (b). The values of
the angles
are in excellent agreement with the experimental ones: 
$sin(\theta) = 0.73$ and $sin(\theta) = 0.43$ for a Mach number 
$13$ and $24$, see Fig.(\ref{Exp}). 
The GPE simulations also reproduce the fan of
precursors moving upstream of the defect~\cite{footnote1}.
By comparing panels (a) and (b) of Fig.\ref{GPE_sim}, one sees that
the longer the imaging time, the longer the spatial distance the
precursors have traveled.  

\begin{figure}[htbp]
    \begin{center}
\includegraphics[angle=-90,width=4.cm,clip]{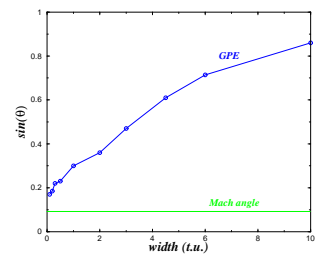} 
      \caption{Conical wave front aperture as a function of
the obstacle width (in ``trap units'', 1 t.u. =3.74 $\mu m$) calculated
with GPE. The green line is the Mach
        angle.}
  \label{aw}
    \end{center}
  \end{figure}

Both the spatial period of the precursors, and the aperture of the
cone decrease for an increasing Mach number $v/c_s$, but this latter is
always significantly larger than the Mach angle such that 
$\sin(\theta)= v/c_s$.
As the flow velocity of the expanding BEC is not uniform, but rather
radial, the geometric angle formed by the tangent of the obstacle
through the BEC center and the line joining the obstacle and the BEC
centers has to be added to the Mach angle.
In addition to this, the geometrical shadow effect provides a simple,
ballistic, explanation also for the strong density reduction inside
the cone. 
This interpretation has been verified by repeating the simulations
for obstacles of smaller size: as it is shown in
Fig.\ref{GPE_sim}(cd), the aperture of the cone significantly
decreases and asymptotically approaches the Mach angle
in the limit of small defect, see Fig.(\ref{aw}).

%Comparing the result of our simulations with recent related work~\cite{el06},
%several differences have to be noted. 
%The precursors were absent, and the low density area inside the cone
%was replaced by a fan of solitons. This latter issue can be explained
%by the different geometry of the flow, while the precursors will be
%shown in the following to be a general consequence of matter-wave
%interference between incoming BEC and atoms which have been scattered
%off the obstacle. 

{\it Bogoliubov-\u Cerenkov theory}.
A simple way of getting a qualitative physical understanding of the
experimental and numerical results is to use an approximate theory where
the condensate is assumed to be homogeneous and uniformly flowing at a
constant speed $\vv_0$ and the defect, localized around $\xx=0$,
induces a weak density
perturbation~\cite{Polar_superfl,astrakharchik04,ianeselli}.  
Under this approximations, analytical results for the stationary
solution of the GPE can be obtained by means of the Bogoliubov
linear response theory. 
As boundary condition, an incident plane wave of the form
$\psi_0(\xx,t)=\psi_0\,e^{i(\kk_0\xx-\omega_0 t)}$ is taken, where we
have set $\hbar\kk_0=m\vv_0$, $\rho_0=|\psi_0|^2$,
$\hbar\omega_0=m\vv_0^2/2+g\rho_0$. Here $g=4\pi\hbar^2 a_s/m$ in
terms of the atomic $s$-wave scattering length $a_s$ and mass $m$. The
sound velocity is $c_s=\sqrt{g\rho_0/m}$. 

\begin{figure}[htbp]
    \begin{center}
\includegraphics[width=8cm,clip]{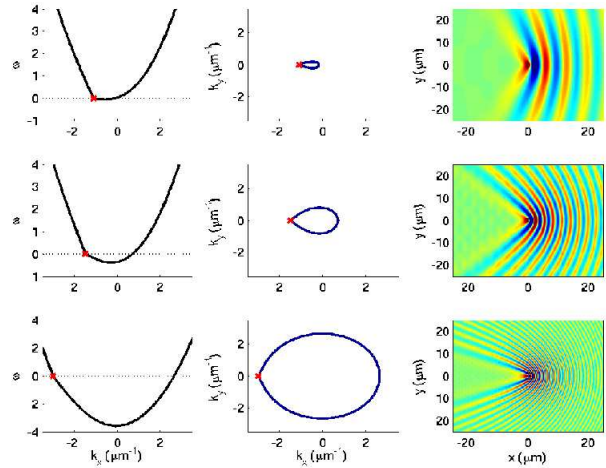}
      \caption{Dispersion of Bogoliubov modes (left
  column), curve $\Gamma$ in $\kk$ space describing the
  resonantly excited Bogoliubov modes (center
  column) and spatial density profile (right colomn). The supersonic
  flows are  $v_0/c_s=1.1, 1.5, 3$ in each row, respectively.
      \label{fig:Supersonic}}
    \end{center}
  \end{figure}

The BEC wavefunction
$\psi(\xx,t)=\big[\psi_0(\xx)+\delta\psi(\xx,t)\big]\,e^{-i\omega_0 t}$ 
is the sum of the unperturbed wavefunction $\psi_0(\xx,t)$ plus the
small perturbation $\delta\psi(\xx,t)$.
Introducing the compact notation $\delta{\vec
  \psi}(\xx,t)=(\delta\psi(\xx,t),\delta\psi^*(\xx,t) )^T$,
%\begin{equation}
%  \label{eq:displ}
%\delta{\vec \psi}(\xx,t)=
%\left(
%\begin{array}{c}
%\delta\psi(\xx,t) \\
%\delta\psi^*(\xx,t)
%\end{array}
%\right),
%\end{equation}
the evolution can be written in the form of a linear equation~\cite{Polar_superfl}:
\begin{equation}
  \label{eq:Bogo}
  i\hbar\frac{\partial}{\partial t}\delta{\vec \psi}={\mathcal L}\cdot \delta{\vec
  \psi}+{\vec F}_d,
\end{equation}
with a source ${\vec F}_d$ proportional to the defect potential
$V_d$: 
\begin{equation}
  \label{eq:F_d}
{\vec F}_d(\xx)=
V_d(\xx)\,\left(
\begin{array}{c}
\psi_0(\xx) \\
-\psi_0^*(\xx)
\end{array}
\right)
. 
\end{equation}
The Bogoliubov matrix ${\mathcal L}$ has the usual form:
\begin{equation}
  \label{eq:BogoL}
{\mathcal L}=
\left(
\begin{array}{cc}
-\frac{\hbar^2}{2m}\nabla^2+\rho_0 g & \rho_0 g \,e^{2i\kk_0 \xx} \\
-\rho_0 g \,e^{-2i\kk_0 \xx} & -\left[-\frac{\hbar^2}{2m}\nabla^2+\rho_0 g\right]
\end{array}
\right)\;.
\end{equation}
and its eigenvalues give the energies of the elementary excitations of 
the system around the unperturbed stationary state. In our specific
case of a homogeneous and homogeneously flowing BEC, the Bogoliubov
dispersion law is
\begin{equation}
  \label{eq:BogoEnerg}
  \omega(\kk)=\vv\cdot(\kk-\kk_0)\pm
\sqrt{\frac{\hbar(\kk-\kk_0)^2}{2m}
\Big(\frac{\hbar(\kk-\kk_0)^2}{2m}+2g\rho_0\Big)}
\end{equation}
As we can see in the dispersion spectra shown in the left column
of Fig.\ref{fig:Supersonic}, the main effect of the flow consists of the
additional term $\vv_0\cdot(\kk-\kk_0)$ which tilts the dispersion
and add $\vv_0$ to the propagation group velocity of all the
Bogoliubov excitations. 
The $\pm$ branches correspond to respectively the particle- and
  the hole-like branches of the Bogoliubov dispersion, and are images
of 
each other under the transformation $\kk\rightarrow
2\kk_0-\kk$, $\omega\rightarrow-\omega$.
The steady state in the presence of the defect potential $V_d$ is
obtained from the motion equation \eq{Bogo} as:
\begin{equation}
  \label{eq:stationary}
\delta{\vec \psi}_d=-\big({\mathcal L-i\,0^+\,{\mathbf 1}}\big)^{-1}
\cdot {\vec F}_d.
\end{equation}
The infinitesimal imaginary term is required to satisfy the boundary
condition that at $t=-\infty$ no Bogoliubov excitations were present.
As can be easily seen from the resonant denominator, the
time-independent defect potential excites the
Bogoliubov modes whose energy is $\omega(\kk)=0$.
Graphically, these $\kk$ modes can be identified in the plots in the left
column of Fig.\ref{fig:Supersonic} as the intersection points of the
dispersion surface $\omega(\kk)$ with the $\omega=0$ plane.
Modulo a Galilean transformation, this condition corresponds to the
usual \u Cerenkov resonance condition~\cite{jelley}.
Depending on whether the flow speed is slower or faster than the speed
of sound in the BEC, two regimes can be identified.

In the subsonic regime $v_0<c_s$, no intersection exists at
$\kk\neq\kk_0$, which physically means that no Bogoliubov mode
can be resonantly excited by the defect.
The BEC is superfluid and can flow around the
defect without suffering any dissipation, in agreement with the
Landau criterion of superfluidity.

On the other hand, in the supersonic regime, $v_0>c_s$, the set of $\kk$
vectors satisfying $\omega(\kk)=0$ is not empty, but rather
corresponds to the closed curve $\Gamma$ shown in the central column
of Fig.\ref{fig:Supersonic}.
Some of the kinetic energy associated to the flow is therefore
dissipated as radiation of Bogoliubov modes.
In real space, the perturbation radially propagates from the defect with a
velocity fixed by the group velocity of the mode $\vv_g=\nabla_\kk\omega$.  
For each mode $\kk$, the direction of $\vv_g$ corresponds to the
outward normal to the curve $\Gamma$.
Since we are here considering a stationary state, the density
perturbation propagates to infinity in all available
directions. Yet, even though 
$\Gamma$ is a closed curve, the possible directions of the
group velocity do not fill the whole unit sphere: because of the
singularity at $\kk_0$ there exists a region of excluded
directions, which 
in the plots in the left column of 
Fig.\ref{fig:Supersonic} correspond to the unperturbed region inside the \u 
Cerenkov cone.

The spatial profiles of the density perturbation predicted by the
linear theory can be understood by analyzing the different regions of
$\Gamma$.  
Close to the singular point at $\kk=\kk_0$, the curve consists of two 
straight lines  separated by an angle $2\phi$ defined by the usual
\u Cerenkov condition $\cos\phi=c_s/v_0$. 
This region corresponds to the small wavevector part of the Bogoliubov
spectrum, which is indeed linear and allows for an interpretation of
the corresponding radiation in terms of the usual \u Cerenkov effect
in non-dispersive media.
In the plots in the right column of fig.\ref{fig:Supersonic}, the
strongest density perturbation lies on the Mach cone of aperture
$\theta$ 
such that $\sin\theta=c_s/v_0$, and directed in the downstream
direction; being the flow 
velocity larger than the sound velocity, the density
perturbation is dragged away from the defect.
The rounded region of $\Gamma$ for $\kk$ nearly anti-parallel to $\kk_0$
is absent in the standard theory of the \u Cerenkov effect in
non-dispersive media, and originates from the quadratic,
single-particle-like, part of the Bogoliubov spectrum. 
The group velocity at the right-most point of $\Gamma$ is equal to
$v_g=\big[c_s^2/v_0-v_0\big]$ so that the perturbation propagates in
the upstream direction with respect to the BEC flow and gives an
oscillating density modulation of wave-vector $|\kk-\kk_0|=2 m
(v_0^2-c_s^2)^{1/2} / \hbar$. 
This is a direct consequence of matter-wave interference between the
incident BEC and the scattered wave off the obstacle~\cite{footnote}.
In the intermediate points on the curve, there is a one-to-one correspondence
between excited Bogoliubov modes of wavevector $\kk$ and the direction 
  ${\hat \xx}$  such that ${\hat \xx}$ is parallel to the group
velocity  $\vv_g(\kk)$. 
For each point $\xx=x {\hat \xx}$ in space, this correspondence
determines the wavevector $\kk$ of the Bogoliubov mode that is able to
reach 
it, and consequently the wavevector $\kk_m=\kk_0-\kk$ of the density
modulation in its vicinity. 
Despite the underlying approximations, the features of the Bogoliubov
theory qualitatively reproduce the full GPE simulations and therefore
clarify the interpretation in terms of \u Cerenkov waves of the 
experimental findings.

In conclusion, we have presented a combined theoretical and experimental study
of the response of a Bose-Einstein condensate flowing against
a localized obstacle at supersonic velocity~\cite{Gladish}.
A numerical study of the full Gross-Pitaevskii equation has been
performed to fully simulate the experiment.
The findings are in agreement with the experiment, and have been
interpreted within the Bogoliubov approximation  
in terms of the \u Cerenkov emission of phonons by the defect:
this creates a Mach cone density modulation downstream of the defect,
as well as an additional upstream fan-shape perturbation, which
originates from the interference between the particle-like Bogoliubov
excitations and the underlying BEC. 
For growing strengths and sizes of the obstacle, the numerical study
shows a continuous evolution of the density modulation towards a
soliton train~\cite{El}, as typical of the dynamical evolution  of
shock-waves in dispersive, dissipationless nonlinear 1D wave equations 
\cite{shock}. 

{\it Acknowledgments.}
We are indebted with E.A. Cornell and P. Engels for providing us the
experimental data shown in Fig.(\ref{Exp}). AS gratefully thanks
P. Engels for several clarifications on the experimental results
and P. Krapivsky for discussions on the formation of 
shock waves in nonlinear systems. IC is grateful to C. Ciuti for continuous discussions
on the \u Cerenkov effect in quantum 
fluids. We also aknowledge a collaboration with C. Menotti 
at an early stage of the work.
Work partially supported by the U.S. Department of Energy, contract
DE-AC52-06NA-25396.


\begin{thebibliography}{99}
\bibitem{footnote1} Differently from~\cite{jackson98}, no evidence for
  the shedding of vortices was found in our simulations.
\end{thebibliography}
\end{document}